\documentclass[fleqn,10pt]{wlscirep}
\usepackage{amsmath,amssymb,amsfonts,graphicx,bm,dcolumn,color,mathrsfs,textcomp,lmodern}
\usepackage[utf8]{inputenc}
\usepackage[T1]{fontenc}
\title{Saturation of radiative heat transfer due to many-body thermalization}

\author[1,2,*]{Ivan Latella}
\author[2]{Riccardo Messina}
\author[3]{Svend-Age Biehs}
\author[1]{J. Miguel Rubi}
\author[2,*]{Philippe Ben-Abdallah}
\affil[1]{Departament de Física de la Matèria Condensada, Universitat de Barcelona, Martí i Franquès 1, 08028 Barcelona, Spain}
\affil[2]{Laboratoire Charles Fabry, UMR 8501, Institut d'Optique, CNRS, Universit\'{e} Paris-Saclay, 2 Avenue Augustin Fresnel, 91127 Palaiseau Cedex, France}
\affil[3]{Institut f\"{u}r Physik, Carl von Ossietzky Universit\"{a}t, D-26111 Oldenburg, Germany}
\affil[*]{ilatella@ub.edu;pba@institutoptique.fr}

\begin{abstract}
Radiative heat transfer between two bodies saturates at very short separation distances due to the nonlocal optical response of the materials.
In this work, we show that the presence of radiative interactions with a third body or external bath can also induce a saturation of the heat transfer, even at separation distances for which the optical response of the materials is purely local. We demonstrate that this saturation mechanism is a direct consequence of a thermalization process resulting from many-body interactions in the system. This effect could have an important impact in the field of nanoscale thermal management of complex systems and in the interpretation of measured signals in thermal metrology at the nanoscale. 
\end{abstract}

\begin{document}

\flushbottom
\maketitle

\thispagestyle{empty}

\section*{Introduction}

The theory of radiative heat transfer~\cite{PoldervH,LoomisPRB94,JoulainSurfSciRep05,VolokitinRMP07,Song,Cuevas} predicts a divergence of the heat flux exchanged between two bodies kept at constant temperatures as the separation distance $d$ between them tends to zero. During the last decade, theoretical results~\cite{VolokitinPRB01,KittelPRL05,HenkelApplPhysB06,ChapuisPRB08,JoulainJQSRT,SingerJQSRT15} have questioned this divergence and shown that it disappears when a nonlocal optical response~\cite{FordPR84} of the materials is taken into account.
Recently, it has been shown that the divergence of the heat transfer can also be removed at subnanometric separation distances because of the interplay of conductive and radiative heat transfer inside the interacting bodies, which lead to the generation of temperature gradients and in turn to a saturation of the heat flux~\cite{Messina-PRB2016-1,Messina-PRB2016-2}. This effect is, however, limited to small separation distances at which new channels for heat transfer (due to phonon tunneling~\cite{Budaev,Ezzahri,ChiloyanNatComm15,Pendry1,Fong} or electron tunneling~\cite{Messina-Arxiv}) start to play a significant role. The divergence is ultimately removed because thermal equilibrium between the bodies is established at contact~\cite{Pan1,Mulet1,Pan2}.

In all these works, the interacting objects are assumed to be isolated from the environment or from other radiative sources. Here we revisite the near-field heat transfer problem between two solids when a third source of thermal radiation participate to the transfer. This situation is fundamentally different from the usual two-body description because many-body interactions are at work. Several problems in the many-body framework have recently been considered~\cite{PBA-APL2006,PBA-PRB2008,PBA-PRL2011,Zheng,KrugerPRB12,BenAbdallahPRL13,NikbakhtJAP14,Ordonnez,Messina-PRB2016,ZhuPRL16,NikbakhtPRB17,Krugger,LatellaPRB17,PBA-PRL2012,Messina-PRB2013,PBA-PRL2014-1,Messina-PRA2014,PBA-PRL2014-2,Dyakov,LatellaPRAppl15,PBA-PRL2016,Ordonez2,Tervo,LatellaPRB18,He19,Kan19,Czapla,LatellaPRAppied19} and new thermophysical effects have been highlighted.

\begin{figure}
\centering
\includegraphics[scale=1]{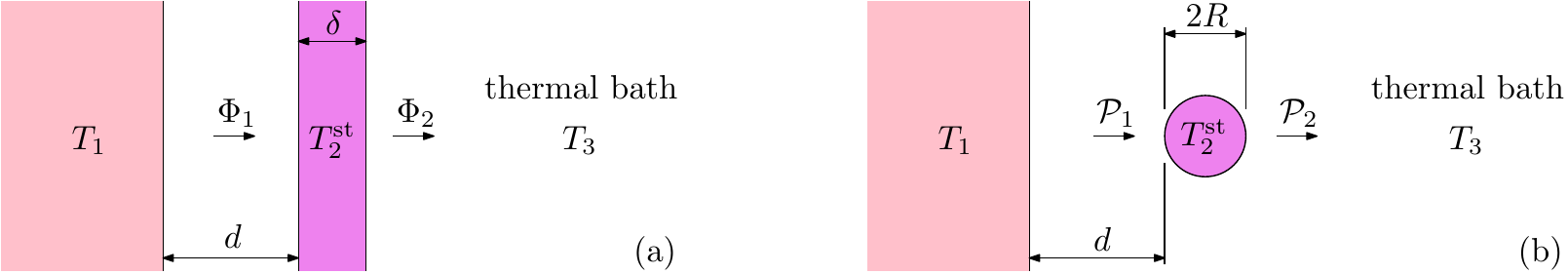}
\caption{Sketch of the system. (a) A membrane is placed close to a substrate at a separation distance $ d $. The substrate is thermalized at a fixed temperature $ T_1 $ and the structure is immersed in an environmental bath of thermal radiation at temperature $ T_3 $. The temperature $ T_2 $ of the membrane is free to reach a steady-state value $ T_2 = T_2^\text{st}$, for which the body achieves heat-transfer equilibrium. (b) A small particle is considered instead of the membrane.}
\label{fig1}
\end{figure}

In this article, we investigate a heat transfer saturation mechanism due to thermalization in many-body systems under nonequilibrium conditions in the absence of nonlocal effects.
The simplest configuration in which such a saturation mechanism can be observed is a two-body system interacting with a thermal bath. 
In order to describe this effect, we consider the heat transfer in the following two simple systems that may mimic many practical situations. The first one is a thin film (i.e. a membrane) that interacts with both a substrate on one side and a thermal bath on the other side, as sketched in Fig.~\ref{fig1}(a). The second system is a small particle which also interacts with both a substrate and an external bath, as represented in Fig.~\ref{fig1}(b). 

\section*{Saturation mechanism for a membrane close to a substrate}

Here we consider a substrate that we denote as body 1 and a membrane of thickness $\delta $, denoted as body 2, separated by a distance $ d $ from body 1. The substrate is thermalized at a fixed temperature $T_1$ and the system interacts with a thermal bath of radiation at temperature $T_3<T_1$, see Fig.~\ref{fig1}(a). The thermal bath acts as a third body in this configuration. The temperature $T_2$ of the membrane is not fixed by a thermostat, so that this body can reach heat-transfer equilibrium at a stationary temperature $T_2=T_2^\text{st}$ for which the net energy flux on the membrane vanishes.

The radiative heat transfer originates from the electromagnetic field produced by the random thermal motion of charges inside the materials~\cite{PoldervH,LoomisPRB94,JoulainSurfSciRep05,VolokitinRMP07,Song,Cuevas}.
Expanding the electromagnetic field in plane-wave components characterized by frequency $\omega$, parallel wave vector $k$, and polarization $p =\mathrm{TM},\mathrm{TE}$, the energy flux (normal component of the Poynting vector) in the different vacuum regions of the system can be written as
\begin{equation}
\Phi_\gamma = \int_0^\infty \frac{ d \omega }{ 2 \pi } \int_0^\infty \frac{ d k }{ 2\pi }\,k \sum_p \sum_{j=1}^2 \hbar \omega\, n_{ j, j+1 } \hat{ \mathcal{ T } }^j_\gamma,
\label{energy_flux}
\end{equation}
where $ \gamma = 1 $ indicates the region between bodies 1 and 2, $ \gamma = 2 $ labels the region on the right of body 2 [see Fig.~\ref{fig1}(a)], and $ n_{ \ell , j } \equiv n_\ell - n_j $ with the thermal distribution function $ n_j = 1 / \big( e^{ \hbar\omega / k_B T_j } - 1 \big) $, $ k_B $ being the Boltzmann constant and $ \hbar $ the reduced Planck constant. Here $ \hat{ \mathcal{ T } }^j_\gamma = \hat{ \mathcal{ T } }^j_\gamma ( k,\omega,p ) $ are the associated energy transmission coefficients given by~\cite{LatellaPRB17}
\begin{equation}
\begin{split}
\hat{ \mathcal{ T } }^1_1 & = \Pi^{\text{pw}} \frac{  \big( 1 -| \rho_1 |^2 \big) \big( 1 -| \rho_2 |^2 \big) }
{ \big| 1- \rho_1 \rho_2 e^{ i 2 k_z d } \big|^2 }
 +  
 \Pi^{ \text{ew} } \frac{ 4 \text{Im} ( \rho_1 ) \text{Im}( \rho_2 ) e^{ - 2 \text{Im} ( k_z ) d } }
{ \big| 1 - \rho_1 \rho_2 e^{ - 2 \text{Im} ( k_z ) d } \big|^2 }, \\   
\hat{ \mathcal{ T } }^1_2 &= \Pi^{ \text{pw} } \frac{  | \tau_2 |^2 \big( 1 - |\rho_1 |^2 \big) }
{ \big| 1 - \rho_1 \rho_2 e^{ i 2 k_z d } \big|^2 },\\
\hat{ \mathcal{ T } }^2_2 & = \Pi^{ \text{pw} } \big( 1 - | \rho_{12} |^2 \big),
\end{split}
\label{trans_coeff}
\end{equation} 
where $ \rho_j = \rho_j ( k,\omega,p ) $ and $ \tau_j = \tau_j ( k,\omega,p ) $ are the optical reflection and transmission coefficients of body $ j $, respectively,
$\rho_{ 12 } = \rho_2 +  ( \tau_2 )^2 \rho_1 e^{ 2 i k_z d }
/\left( 1 - \rho_1 \rho_2 e^{ 2 i k_z d } \right)$
is the reflection coefficient of bodies 1 and 2 together, 
$k_z = \sqrt{ \omega^2/ c^2  - k ^2 } $
is the component of the wave vector perpendicular to the surfaces in the vacuum regions, and the projectors on the propagating and evanescent wave sectors are defined by $ \Pi^\text{pw} \equiv \theta( \omega - c k ) $ and $ \Pi^\text{ew} \equiv \theta ( c k - \omega ) $, respectively, $ c $ being the speed of light in vacuum and $\theta ( x ) $ the Heaviside step function. The coefficients $\rho_j$ and $\tau_j$ depend on the Fresnel reflection coefficients of the interfaces $r_j^p=r_j^p(k,\omega)$ as detailed in Methods; in particular, since the substrate is assumed to be a semi-infinite, dissipative body, we have $\rho_1 ( k,\omega,p ) = r_1^p ( k,\omega )$ and $\tau_1 ( k,\omega,p ) = 0$, which have been used to obtain Eqs.~(\ref{trans_coeff}). Moreover, these transmission coefficients satisfy~\cite{LatellaPRB17} $\hat{ \mathcal{ T } }^j_\gamma = \hat{ \mathcal{ T } }^\gamma_j $, from which $ \hat{ \mathcal{ T } }^2_1 = \hat{ \mathcal{ T } }^1_2 $.

\begin{figure}[!t]
\centering
\includegraphics[scale=1]{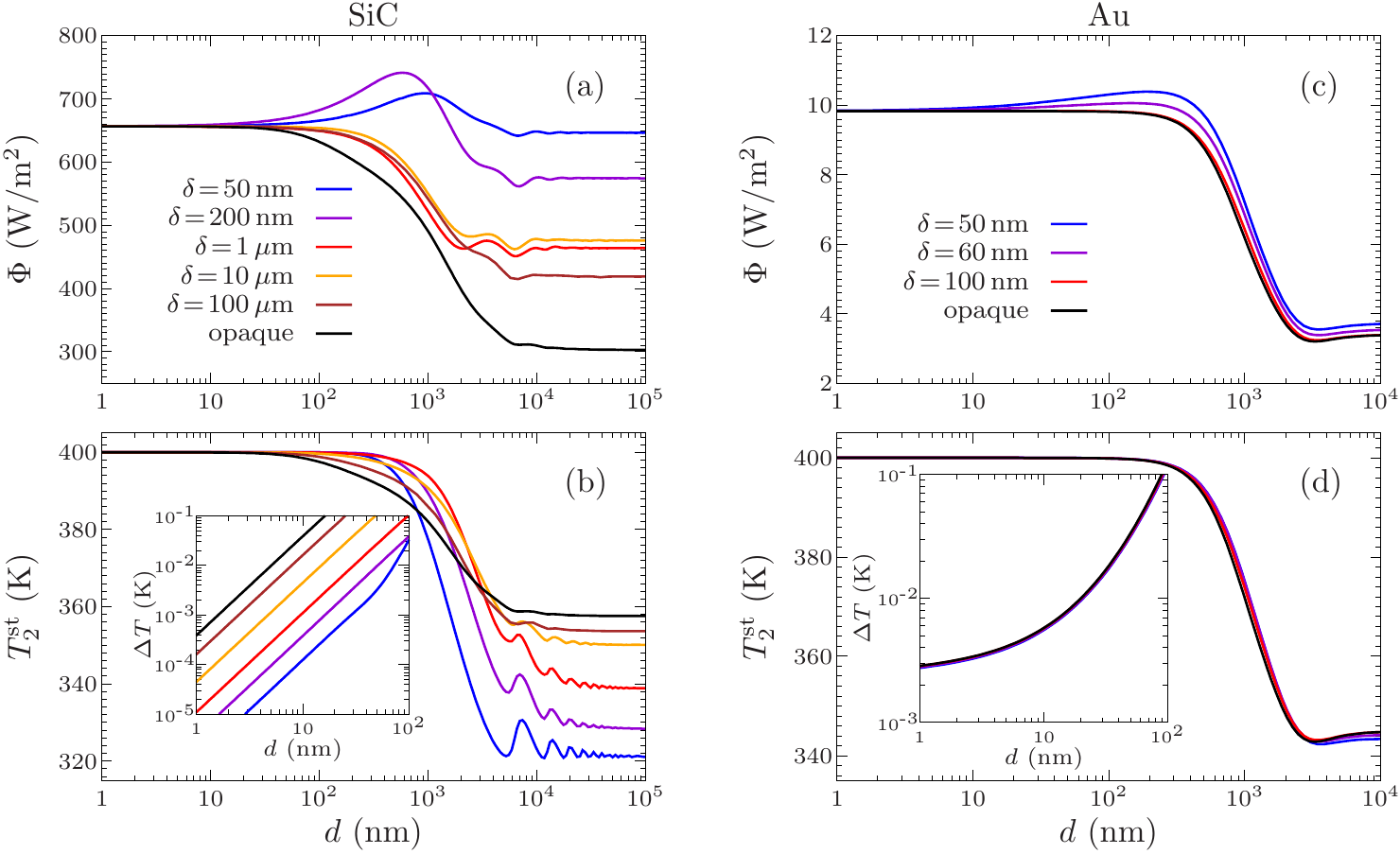}
\caption{Energy flux and steady-state membrane temperature as a function of the separation distance. The substrate and membrane are made of SiC in (a) and (b), while (c) and (d) correspond to Au. Here the substrate and bath temperatures are $ T_1 = 400\,$K and $ T_3 = 300\,$K, respectively, and the results are shown for several values of the membrane thickness $\delta$. The insets show the behavior of $ \Delta T = T_1-T_2^\text{st} $ at small $ d $ (all curves approximately  coincide for Au).}
\label{fig2}
\end{figure}

The steady-state temperature $ T_2^\mathrm{ st } $ of the membrane is obtained by requiring a vanishing net energy flux on this body, so that
$\Phi_1 ( T_2^\mathrm{ st } ) - \Phi_2 ( T_2^\mathrm{ st } ) = 0$.
Since the fluxes depend on the separation distance, the steady-state temperature depend on $ d $ as well. Hereafter we solve this equation by taking $ T_1 = 400\,$K and $T_3=300\,$K. In Fig.~\ref{fig2}(a), we show 
$\Phi\equiv \Phi_1(T_2^\mathrm{ st })=\Phi_2( T_2^\mathrm{ st } )$
with respect to the separation distance $ d $ for several values of the thickness $\delta$ in the case in which both the slab and the membrane are made of silicon carbide (SiC).
We observe a saturation of the heat flux at relatively  large separation distances, where nonlocal effects are completely negligible. This saturation mechanism is directly related to the dependence of the temperature difference 
$\Delta T= T_1-T_2^\mathrm{ st } $
on the separation distance $ d $. As shown in the inset of Fig.~\ref{fig2}(b), $ \Delta T $ is proportional to $ d^2$ at short separations. Moreover, since the flux $ \Phi$ approaches a constant at small $ d $, the ratio $\Phi /\Delta T$ (i.e., the heat transfer coefficient) scales as $1/d^2$ in this regime, as already outlined in the literature~\cite{Pendry,MuletMTE02} for polar materials like SiC. Notice that the asymptotic value of $ \Phi $ at short separations does not depend on the width $ \delta $: in the limit $ d \to 0 $, $ \Phi $ correspond to the energy flux radiated to the environment by a single semi-infinite body (see below). 

We now derive analytic expressions for the saturation heat flux and the temperature difference. Setting $ T_2^\mathrm{ st }=T_1-\Delta T$ and assuming small $\Delta T$, at $ T_2 = T_2^\mathrm{ st }$, we have $n_2=n_1-(\partial n_1/\partial T_1) \Delta T$ to leading order in $\Delta T$. 
Taking this into account,  we can rewrite Eq.~\eqref{energy_flux} as
$\Phi_\gamma=a_\gamma \Delta T + b_\gamma$
with
\begin{equation}
a_\gamma = \int_0^\infty \frac{ d \omega }{ 2\pi } \hbar\omega \frac{ \partial n_1 }{ \partial T_1 } \int_0^\infty \frac{ d k }{ 2\pi }\,k \sum_p \left( \hat{ \mathcal{ T } }^1_\gamma - \hat{ \mathcal{ T } }^2_\gamma  \right), \qquad
b_\gamma =  \int_0^\infty \frac{ d \omega }{ 2\pi } \hbar\omega n_{1,3} \int_0^\infty \frac{d k }{ 2\pi }\, k \sum_p \hat{ \mathcal{ T } }^2_\gamma. 
\label{abgamma}
\end{equation}
Then, the equilibrium condition $\Phi_1 ( T_2^\mathrm{ st } ) = \Phi_2 ( T_2^\mathrm{ st } ) $ leads to
\begin{equation}
\Phi =\frac{a_1 b_2-a_2b_1}{a_1-a_2}, \qquad
\Delta T = \frac{b_2 -b_1}{a_1-a_2}. 
\label{Phi_Delta}
\end{equation}

Let us now consider how this saturation mechanism is modified for metallic materials. At short separation distances, it is well known that the heat transfer between metals is first dominated by the TE-polarization contribution  and the  $1/d^2$ behavior associated to TM waves is usually recovered at subnanometer separation distances. However, this divergence disappear because of nonlocal effects~\cite{ChapuisPRB08}. We show below that in metallic many-body systems a saturation of heat flux can exist at larger separation distances. To this aim, we consider a system made of a gold (Au) membrane suspended above a Au substrate. 

The results for the heat flux and the steady-state temperature for a slab and a membrane made of Au are shown in Fig.~\ref{fig2}(c) and Fig.~\ref{fig2}(d), respectively. A saturation of the heat flux at short separations is observed also for this material. We highlight that the dependence on the thickness of the membrane is weak for $\delta$ larger than $ 100\,$nm. This is due to the fact that the electromagnetic field is completely screened for such thicknesses and the membrane becomes practically opaque. Furthermore, the behavior of $ \Delta T $ in this case is shown in the inset of Fig.~\ref{fig2}(d). Clearly, the temperature difference is not proportional to $ d ^2 $ at small $ d $ because of the contribution of TE-polarized waves (TM polarization dominates well below the nanometer scale). Such a behaviour emphasizes that here the saturation mechanism is different from that for polar materials.

\subsection*{Asymptotic short-distance behavior}

Here we analyze the asymptotic behavior of the heat flux and temperature difference at short distances. We discuss separately TM and TE polarizations, the former being the dominant contribution for the considered polar material and the latter for the metal. By neglecting the contribution of propagating waves at close separation distances, the coefficient $ a_1 $ defined in expression~(\ref{abgamma}) reads
\begin{equation}
a_1 = \sum_p \int_0^\infty \frac{ d \omega }{ 2\pi } \hbar\omega \frac{ \partial n_1 }{ \partial T_1 }
\int_{ \omega/c }^\infty \frac{ d k }{ 2\pi }\, k \frac{ 4 \text{Im} ( \rho_1 )
\text{Im} ( \rho_2 ) e^{ - 2 \text{Im} ( k_z ) d } }
{ \big| 1 - \rho_1 \rho_2 e^{ - 2 \text{Im} ( k_z ) d } \big|^2 }.
\label{eq_app1}
\end{equation}
We also assume that the thickness of the membrane is large as compared with the separation distance, which corresponds to the limit $ \delta \rightarrow\infty$ in the expressions for the reflection coefficients and therefore, $\rho_2 ( k,\omega,p ) \to r_2^p ( k,\omega )$. For simplicity, we consider that the materials are identical and thus define $r^p\equiv r_1^p=r^p_2$ and $\varepsilon(\omega)\equiv \varepsilon_1(\omega) = \varepsilon_2(\omega)$.

\subsubsection*{Polar materials}

The heat exchange between polar materials at short separations can be studied in the electrostatic limit. In this limit, only large wavevectors $k\gg k_0$ contribute to the heat exchange, where $k_0=\omega/c$, and the normal component $k_z$ can be approximated by $i k$. Moreover, the Fresnel reflection coefficient for TM polarization takes the form
$r^\text{TM} \simeq ( \varepsilon -1 )/( \varepsilon +1 )\equiv r $, while for TE-polarized waves this coefficeint vanishes as $ r^\text{TE} \simeq \frac{1}{4}( \varepsilon - 1 )(k_0/k)^2$.
Keeping only the TM polarization, Eq.~(\ref{eq_app1}) can then be written as~\cite{Rousseau}
\begin{equation}
a_1 =\frac{1}{d^2}
\int_0^\infty \frac{ d \omega }{ 2\pi } \hbar\omega \frac{ \partial n_1 }{ \partial T_1 } 
\frac{4\text{Im}^2 ( r ) } { \text{Im} ( r^2 )} 
\text{Im} \int_{ 0 }^\infty \frac{ d x }{ 2\pi }\, x 
\frac{   r^2   e^{ - 2 x } }
{  1 - r^2 e^{ - 2 x } }
=\frac{1}{ d^2}
\int_0^\infty \frac{ d \omega }{ 2\pi } \hbar\omega \frac{ \partial n_1 }{ \partial T_1 } 
\frac{\text{Im}^2 ( r ) \text{Im} \left[ \text{Li}_2(r^2) \right] } { 2\pi \text{Im} ( r^2 )} ,
\end{equation}
where $\text{Li}_2(w)$ denotes the dilogarithm function.
Hence, in the limit $ d \to 0 $, the coefficients $ a_2 $, $ b_1 $, and $ b_2 $ in Eqs.~(\ref{Phi_Delta}) remain finite since they have propagating waves contribution only, while $ a_1 $ diverges as $ d^{-2} $ because of the contribution of evanescent waves in TM polarization. Thus, $ \Delta T \to 0 $ and $ \Phi \to b_2 $  as $ d \to 0 $. More explicitly, in this limit the energy flux becomes
\begin{equation}
\Phi = \int_0^\infty \frac{ d \omega }{ 2\pi } \hbar\omega n_{ 1,3 } \int_0^{ \omega/c } \frac{ d k }{ 2\pi }\, k \sum_p \big( 1 - |\rho_{12}|^2 \big),
\label{Phi_limit}
\end{equation}
which corresponds, as anticipated above, to the heat exchanged between bodies 1 and 2 together at temperature $ T_1 $ and an environment at temperature $ T_3 $. Notice that when bodies 1 and 2 are made of the same material, so that $ r_2^p ( k,\omega ) = r_1^p ( k,\omega ) $, one has
$ \rho_{12}( k,\omega,p )  = r_1^p ( k,\omega )$ in the limit $d \to 0$.

\subsubsection*{Metals}

For metals close to room temperature, the heat flux in the electrostatic limit is dominated by TM polarization at subnanometer separation distances and the traditional $1/d^2$ divergence is regularized by the presence of nonlocal effects~\cite{VolokitinPRB01,ChapuisPRB08}. However, at these separation distances, other mechanisms superimpose to the radiative transfer such as phonon~\cite{Budaev,Ezzahri,ChiloyanNatComm15,Pendry1,Fong} or electron tunneling~\cite{Messina-Arxiv}. Close to contact, these channels even dominate the heat transfer. For separation distances slightly larger (usually for $d\gtrsim 1\,$nm), the radiative transfer in metals is entirely driven by TE-polarization states and nonlocal optical effects~\cite{VolokitinPRB01,ChapuisPRB08} do not play any role. In this case, the imaginary part of $r^\mathrm{TE}$ decays with respect to  $ k $, so that the flux saturates~\cite{ChapuisPRB08} for a wave vector $k_\mathrm{max}=\omega_p/c$ before increasing again close to contact, where $\omega_p$ is the plasma frequency of the metal (see Methods). Typically, this saturation is observable between $d\sim 1\,$nm and separation distances similar to the skin depth of the metal evaluated at $\omega_p$ (about $20\,$nm for Au). Nonetheless, this effect takes place at separation distances which are one order of magnitude smaller than the saturation distance induced by the thermalization process, as shown in Fig.~\ref{fig2}(c) in our example for Au.
In addition, since the transport is mediated by TE-polarized waves, the heat-transfer coefficient $a_1$ given by Eq.~(\ref{eq_app1}) remains finite at short separations.
Although $a_1$ is finite in this regime, it is large as compared with $a_2$ in Eq.~(\ref{Phi_Delta}) because the latter only accounts for the contribution of propagating waves. Thus, from Eq.~(\ref{Phi_Delta}), one obtains $ \Phi \approx b_2 $ and therefore the flux is approximately given by Eq.~(\ref{Phi_limit}), while $\Delta T \approx (b_2 - b_1)/a_1$, which is small but finite in the considered limit. This behaviour is observed in the inset of Fig.~\ref{fig2}(d) for Au.

\subsection*{Opaque membrane}

In the example of the heat flux saturation for the metal, we have shown that the results are not sensitive to the thickness of the membrane when this is larger than 100\,nm. This is due to the fact that, because of dissipation, the electromagnetic field is completely screened inside the material. In other words, the membrane becomes opaque when it is thick enough. Such a screening occurs also in polar materials, but for thicknesses typically larger than for metals: In our example of SiC, the opaque-membrane limit takes place at $ \delta $ much larger than $ 100\,\mu$m.

Assuming that the membrane is opaque introduces a simplification in the heat-transfer problem, which then can be described through next-neighbor interactions only. This can be seen by noting that the factor $ e^{ i k_{ z  2 } \delta } $ in the optical reflection and transmission coefficients vanishes for large $ \delta $ (see Methods), that is, when the membrane is opaque, because $ \text{Im}(k_{ z  2 }) > 0 $ for dissipative materials. Under these conditions, we have $ \rho_2 ( k,\omega,p )  \to r_2^p ( k,\omega ) $ and $ \tau_2 ( k,\omega,p ) \to 0 $,
and therefore, in this opaque-membrane limit the energy transmission coefficients (\ref{trans_coeff}) become
\begin{equation}
\hat{ \mathcal{ T } }^1_1 = \frac{ \Pi^{\text{pw}} \big( 1 -| r^p_1 |^2 \big) \big( 1 -| r^p_2 |^2 \big) }
{ \big| 1- r^p_1 r^p_2 e^{ i 2 k_z d } \big|^2 } 
 +  
\frac{ \Pi^{ \text{ew} } 4 \text{Im} ( r^p_1 ) \text{Im}( r^p_2 ) e^{ - 2 \text{Im} ( k_z ) d } }
{ \big| 1 - r^p_1 r^p_2 e^{ - 2 \text{Im} ( k_z ) d } \big|^2 }, \qquad   
\hat{ \mathcal{ T } }^2_2 = \Pi^{ \text{pw} } \big( 1 - | r^p_2 |^2 \big),
\label{trans_coeff_opaque}
\end{equation} 
and $\hat{ \mathcal{ T } }^1_2 = \hat{ \mathcal{ T } }^2_1 = 0$. We emphasize that these energy transmission coefficients are expressed in terms of the single-interface reflection coefficients only. The energy flux $\Phi$ and stationary temperature $T^\mathrm{st}_2$ in the opaque membrane limit are shown in Fig.~\ref{fig2} as a function of the separation distance for SiC and Au. 
It can be seen that the value of $\Phi$ for the opaque membrane gives for all distances a lower bound on the steady-state heat flux. Furthermore, we observe that $T^\mathrm{st}_\mathrm{opaque} < T^\mathrm{st}_2$ in the near-field regime and $T^\mathrm{st}_\mathrm{opaque} > T^\mathrm{st}_2$ in the far-field regime.

\section*{Saturation mechanism for a particle close to a substrate}

In the previous section, we have analyzed a mechanism of saturation of the heat exchange in a system with planar geometry. In this section, we extend the discussion to a situation in which a small particle is considered instead of a membrane. The particle is assumed small as compared with the thermal wave length, so that it can be modeled as a single dipole in the dipolar approximation.

The system thus consists of a substrate at temperature $T_1$ filling the half-space $z<0$, a particle of radius $R$ at temperature $T_2$ centered at the point $ \mathbf{r} = ( x, y, d + R )$, and a radiative thermal bath at temperature $T_3$ surrounding the particle, see Fig.~\ref{fig1}(b). The power absorbed by the particle at the point $\mathbf{r}$ and instant $t$ is given by
\begin{equation}
\mathcal{P}_\mathrm{abs}=\left\langle \frac{d \mathbf{p} ( \mathbf{r}, t ) }{dt} \cdot \mathbf{E} ( \mathbf{r}, t ) \right\rangle
\label{power_abs}
\end{equation}
where $\langle \,\cdots\rangle$ indicates statistical average, $\mathbf{p}( \mathbf{r}, t )$ is the dipole moment of the particle and $\mathbf{E}( \mathbf{r}, t )$ is the local electric field at the point $\mathbf{r}$. Introducing the Fourier components $f(\omega)$ at frequency $\omega$ such that $ f(t) = 2\mathrm{Re} \left[ \int_0^\infty \frac{ d \omega}{ 2\pi } f( \omega ) e^{ - i \omega t}\right]$, Eq.~(\ref{power_abs}) can be written as
$\mathcal{P}_\mathrm{abs} = \int_0^\infty \frac{ d \omega}{ 2\pi } \mathcal{P}(\omega)$,
where the spectral power is given by
\begin{equation}
\mathcal{P}(\omega) 
= \int_0^\infty \frac{ d \omega' }{ 2\pi } 2 \omega 
\mathrm{Im} \left[ \langle \mathbf{p} ( \mathbf{r}, \omega ) \cdot \mathbf{E}^* ( \mathbf{r}, \omega' ) \rangle e^{ - i (\omega - \omega' ) t}\right].
\label{spectral_power}
\end{equation}

The Fourier components of the local field can be separated into the incident field $\mathbf{E}^\mathrm{inc}( \mathbf{r}, \omega )$ and the induced field $\mathbf{E}^\mathrm{ind}( \mathbf{r}, \omega ) = \omega^2 \mu_0 \mathbb{G}( \mathbf{r},\mathbf{r}, \omega ) \mathbf{p}( \mathbf{r}, \omega )$, that is
$\mathbf{E}( \mathbf{r}, \omega ) = \mathbf{E}^\mathrm{inc}( \mathbf{r}, \omega ) + \omega^2 \mu_0 \mathbb{G}( \mathbf{r},\mathbf{r}, \omega ) \mathbf{p}( \mathbf{r}, \omega )$,
where $\mu_0$ is the vacuum permeability and $\mathbb{G}(\mathbf{r},\mathbf{r}',\omega)$ is the dyadic Green's function of the system. The latter can be written as $\mathbb{G}(\mathbf{r},\mathbf{r}',\omega)= \mathbb{G}^{(0)}(\mathbf{r},\mathbf{r}',\omega) + \mathbb{G}^\mathrm{(R)}(\mathbf{r},\mathbf{r}',\omega)$, where the first term is the free space contribution and the second term is the scattering contribution accounting for reflections on the surface of the substrate.
The real part of the free space Green's function is divergent in the coincidence limit $\mathbf{r}' \to \mathbf{r}$, but only its imaginary part contributes to the absorbed power and is given by
$\mathrm{Im} \mathbb{G}^{(0)}( \mathbf{r}, \mathbf{r}, \omega )  = \frac{\omega}{6 \pi c} \mathbb{I}$,
where $\mathbb{I}$ denotes the unit dyad.
The scattering Green's function reads~\cite{Novotny} 
\begin{equation}
\mathbb{G}^\mathrm{(R)}( \mathbf{r}, \mathbf{r}, \omega ) = 
i \int_0^\infty \frac{d k}{ 8\pi} k
\begin{pmatrix}
a & 0 & 0 \\
0 & a & 0 \\
0 & 0 & b 
\end{pmatrix},\qquad
a = \left( \frac{1}{k_z} r_1^\mathrm{TE} - \frac{c^2 k_z}{\omega^2} r_1^\mathrm{TM}  \right) e^{ 2 i k_z ( d + R )}  ,\qquad
b = \frac{2 c^2 k^2}{\omega^2 k_z} r_1^\mathrm{TM} e^{ 2 i k_z ( d + R ) } . 
\label{GR-a-b}
\end{equation}

Furthermore, the dipole moment of the particle can be decomposed into a fluctuating part $\mathbf{p}^\mathrm{fl}( \mathbf{r}, \omega )$ and an induced part resulting from the incident field $\mathbf{E}^\mathrm{inc}$ and the field produced by the dipole itself and then scattered by the surface, i.e. $\omega^2 \mu_0 \mathbb{G}^\mathrm{(R)}( \mathbf{r},\mathbf{r}, \omega ) \mathbf{p}( \mathbf{r}, \omega )$, so that
\begin{equation}\begin{split}
\mathbf{p}( \mathbf{r}, \omega ) 
&= \mathbf{p}^\mathrm{fl}( \mathbf{r}, \omega ) 
+ \varepsilon_0 \alpha (\omega) \mathbf{E}^\mathrm{inc}( \mathbf{r}, \omega )
+ \frac{\omega^2}{c^2} \alpha (\omega) \mathbb{G}^\mathrm{(R)}( \mathbf{r},\mathbf{r}, \omega ) \mathbf{p}( \mathbf{r}, \omega ),
\end{split}\label{dipole_moment}
\end{equation}
where $\varepsilon_0$ is the vacuum permittivity and $\alpha(\omega)$ is the dressed polarizability of the particle (see Methods). Hence, noting that the matrix $\mathbb{G}( \mathbf{r}, \mathbf{r}, \omega ) $ is diagonal, with simple manipulations, the components of the dipolar moment and local field can be written as
\begin{equation}
p_i = \xi_i \Bigl[p^\mathrm{fl}_i + \varepsilon_0\alpha E^\mathrm{inc}_i\Bigr],\qquad
E_i = \xi_i \Bigl[\Bigl(1+\frac{\omega^2}{c^2}\alpha \,G^{(0)}_{ii}\Bigr)E^\mathrm{inc}_i+\omega^2\mu_0G_{ii}p^\mathrm{fl}_i \Bigr],
\label{local_field}
\end{equation}
with
\begin{equation}
\xi_i =\Bigl(1-\frac{\omega^2}{c^2}\alpha\,G^\mathrm{(R)}_{ii}\Bigr)^{-1}.
\label{xi}
\end{equation}
Taking the statistical average and using Eqs.~(\ref{local_field}) leads to
\begin{equation}
\begin{split}
\langle \mathbf{p}( \mathbf{r}, \omega ) \cdot
\mathbf{E}^*( \mathbf{r}, \omega' )  \rangle
&=  \Bigl[\omega'^2 \mu_0 \sum_i 
 \xi_i (\omega) \xi_i^* (\omega') 
G_{ii}^*( \mathbf{r},\mathbf{r}, \omega' ) 
\langle p_i^\mathrm{fl}( \mathbf{r}, \omega )
p_i^\mathrm{fl *}( \mathbf{r}, \omega' ) \rangle \\
&+ 
\varepsilon_0 \alpha (\omega) \sum_i
 \xi_i (\omega) \xi_i^* (\omega') 
\Bigl( 1 +
\frac{\omega'^2}{c^2} \alpha^* (\omega')
G^{(0)*}_{ii}( \mathbf{r},\mathbf{r}, \omega' )
\Bigr)
\langle E_i^\mathrm{inc}( \mathbf{r}, \omega )
E_i^\mathrm{inc *}( \mathbf{r}, \omega' ) \rangle\Bigr] , 
\end{split}
\label{eq_dipole}
\end{equation}
where we have used that the fluctuating part of the dipole moment and incident fields are uncorrelated. To work out this expression, below we first compute the correlations of the incident field.
In what follows, for simplicity, we will omit writing down explicitly the dependence on positions of the fields, Green's functions and correlation matrices, since they will always be evaluated at the same point $\mathbf{r}$ in the coincidence limit.

\begin{figure}[!t]
\centering
\includegraphics[scale=1]{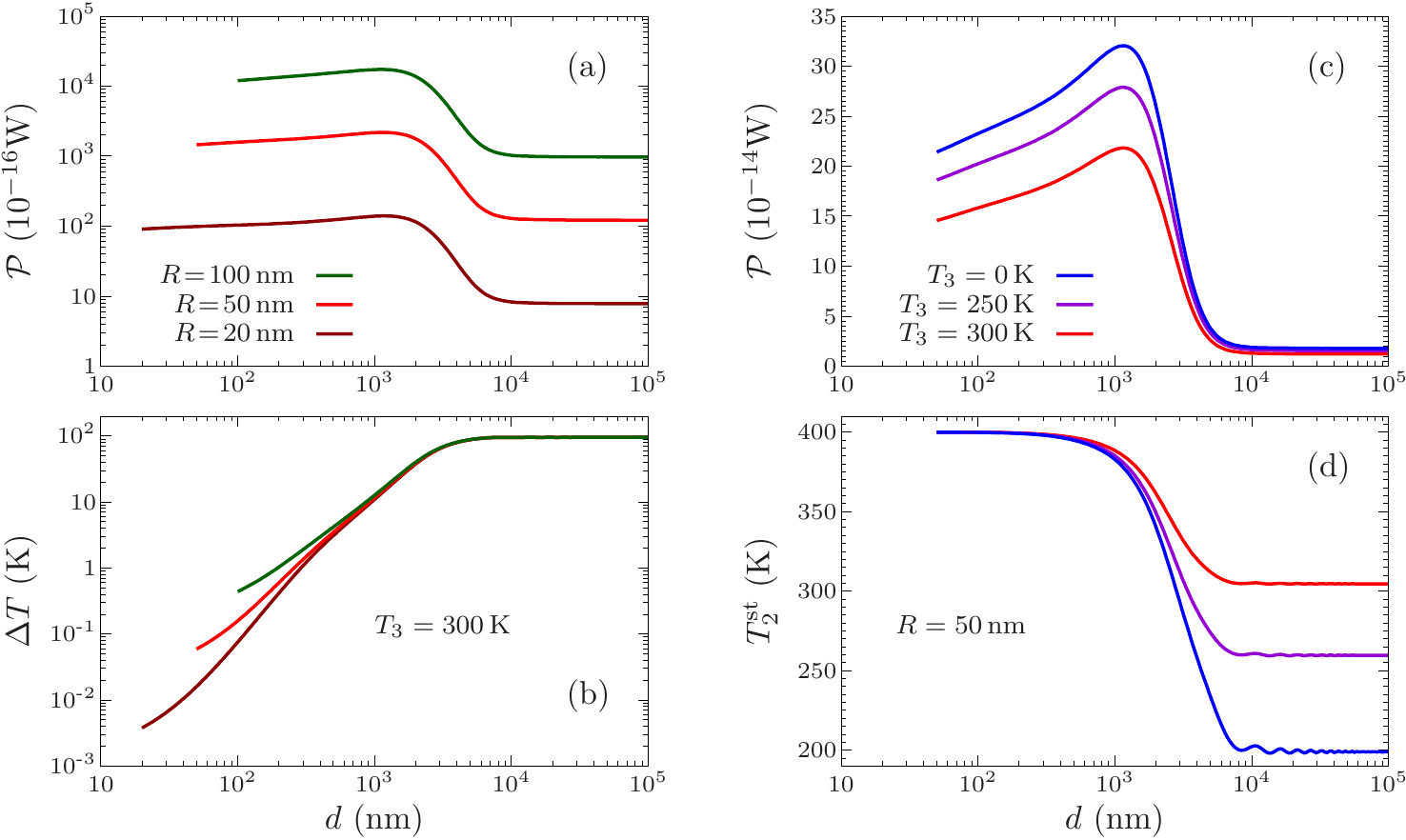}
\caption{Exchanged power and steady-state particle temperature as a function of the separation distance. The substrate and particle are made of SiC. In (a) and (b), the substrate and bath temperatures are $ T_1 = 400\,$K and $ T_3 = 300\,$K, respectively, and the results are shown for several particle radius $R$. In (c) and (d), the substrate and bath temperatures are $ T_1 = 400\,$K and $T_3=0\,$K, $250\,$K, and $300\,$K, respectively, while the particle radius is $R=50\,$nm.}
\label{fig3}
\end{figure}

The incident field can be decomposed into a contribution coming from the substrate $\mathbf{E}^{1}( \omega )$ and a contribution from the bath field $\mathbf{E}^{3}( \omega )$, so that
$\mathbf{E}^\mathrm{inc}( \omega )= \mathbf{E}^{1}( \omega ) + \mathbf{E}^{3}( \omega )$.
The substrate field $\mathbf{E}^{1}( \omega )$ is a direct contribution to the total field at the point $\mathbf{r}$ propagating to the right, while the bath field $\mathbf{E}^{3}( \omega )$ accounts for a direct contribution propagating to the left and a reflected one propagating to the right.
The correlation matrix of the incident field is given by
\begin{equation}
\langle \mathbf{E}^\mathrm{inc}( \omega ) \mathbf{E}^\mathrm{inc \dag }( \omega' ) \rangle 
= \langle \mathbf{E}^{1}( \omega ) \mathbf{E}^{1 \dag }( \omega' ) \rangle 
+\langle \mathbf{E}^{3}( \omega ) \mathbf{E}^{3 \dag }( \omega' ) \rangle ,
\label{correlation_matrix_incident}
\end{equation}
where we have assumed that the substrate and bath fields are uncorrelated.
In addition, when the substrate is in thermal equilibrium with the bath field at, for instance, temperature $T_3$, the correlation matrix of the incident field can be computed from the fluctuation-dissipation theorem
\begin{equation}
\langle \mathbf{E}^\mathrm{inc}( \omega ) \mathbf{E}^{\mathrm{inc} \dag }( \omega' ) \rangle
=\frac{4\pi\hbar\omega^2 }{\varepsilon_0c^2} n_3(\omega) \delta(\omega-\omega') 
\mathrm{ Im } \mathbb{G}( \omega ),
\label{FDT_incident}
\end{equation}
where the total Green's function of the system is used here because, in the absence of the dipole, the incident field is the total field on the right side of the substrate.
Moreover, when evaluated at temperature $T_1$, the correlation matrix of the substrate field can be written as
\begin{equation}
\langle \mathbf{E}^{ 1 }( \omega ) \mathbf{E}^{ 1 \dag}( \omega' ) \rangle
= \frac{ 4 \pi \hbar \omega^2 }{ \varepsilon_0 c^2} n_1(\omega) \delta( \omega - \omega' ) \mathbb{S}( \omega ),
\label{correlation_substrate}
\end{equation}
where we have introduced the matrix $\mathbb{S}(\omega)$ whose explicit form is given in Methods.
The correlation matrix of the bath field can thus be obtained from Eq.~(\ref{correlation_matrix_incident}) using Eq.~(\ref{FDT_incident}) and Eq.~(\ref{correlation_substrate}) evaluated at temperature $T_3$, which gives
\begin{equation}
\langle \mathbf{E}^{ 3 }( \omega ) \mathbf{E}^{ 3 \dag}( \omega' ) \rangle
= \frac{ 4 \pi \hbar \omega^2 }{ \varepsilon_0 c^2} n_3(\omega) \delta( \omega - \omega' ) \mathbb{B}( \omega ),
\label{correlation_bath}
\end{equation}
where
$\mathbb{B}( \omega ) = \mathrm{ Im } \mathbb{G}( \omega ) - \mathbb{S}( \omega )$.
In Methods we also give an explicit expression of the matrix $\mathbb{B}( \omega )$. 
To complete the description of the problem, we need to know the correlation matrix of the fluctuating dipole moment which is given by~\cite{MessinaPRB13} 
\begin{equation}
\langle \mathbf{p}^\mathrm{fl}( \omega ) \mathbf{p}^\mathrm{fl \dag}( \omega' )\rangle 
= 4\pi\hbar \varepsilon_0 n_2(\omega) \delta(\omega-\omega') 
\chi(\omega) \mathbb{I}, 
\label{correlation_dipole}
\end{equation}
where 
$\chi(\omega) = \mathrm{Im} [ \alpha (\omega) ] - \omega^3 | \alpha (\omega) |^2 / (6\pi c^3)$.
By using the correlation functions given above in Eq.~(\ref{eq_dipole}), the spectral power (\ref{spectral_power}) becomes
\begin{equation}
\mathcal{P}(\omega) = \frac{ 4\hbar \omega^3 }{c^2} \chi(\omega)\sum_i |\xi_i(\omega)|^2 
\bigl[ 
n_{1,2}(\omega)S_{ii}(\omega)-n_{2,3}(\omega)B_{ii}(\omega)\bigr] ,
\label{spectral_power_final}
\end{equation}
which manifestly goes to zero at thermal equilibrium.
In view of expression (\ref{spectral_power_final}), the total power absorbed by the particle can be decomposed as $\mathcal{P}_\mathrm{abs} = \mathcal{P}_1 - \mathcal{P}_2$, where
\begin{equation}
\mathcal{P}_{1} = \int_0^\infty \frac{ d\omega }{ 2 \pi } n_{1,2}(\omega) \frac{4\hbar \omega^3}{ c^2 } \chi (\omega)
\sum_i |\xi_i(\omega)|^2 S_{ii} (\omega)
\label{power_substrate_dipole}
\end{equation}
is the power absorbed by the particle due to heat exchange with the substrate in presence of the thermal bath, and
\begin{equation}
\mathcal{P}_{2} = \int_0^\infty \frac{ d\omega }{ 2 \pi } n_{2,3}(\omega) \frac{4\hbar \omega^3}{ c^2 } \chi (\omega)
\sum_i |\xi_i(\omega)|^2 B_{ii} (\omega)
\label{power_bath_dipole}
\end{equation}
is the power emitted by the particle due to its interaction with the bath in presence of the substrate.

We are interested in a situation of heat-transfer equilibrium in which the total power absorbed by the particle vanishes for $T_2=T_2^\mathrm{st}$ at fixed $T_1$ and $T_3$, so that $\mathcal{P}_{1}(T_2^\mathrm{st}) - \mathcal{P}_{2}(T_2^\mathrm{st}) = 0$. This situation is characterized by the stationary temperature of the particle and by the exchanged power
$\mathcal{P}\equiv \mathcal{P}_{1}(T_2^\mathrm{st}) = \mathcal{P}_{2}(T_2^\mathrm{st})$,
which here is studied as a function of the separation $d$. Since we describe the particle in the dipolar approximation, we restrict ourselves to separation distances larger than the radius of the particle, $d>R$ (the distance between the substrate and the center of the particle is thus larger than $2R$). We also emphasize that the approach developed above is appropriate for polar materials, but needs to be suitably modified for metals, introducing the magnetic contribution to the power absorbed by the dipole. With this in mind, here we consider a substrate and a particle made of SiC. In Figs.~\ref{fig3}(a) and \ref{fig3}(b), we plot $\mathcal{P}$ and the associated $ \Delta T = T_1-T_2^\text{st} $, respectively, as a function of the separation $d$ for $T_1=400\,$K and $T_3=300\,$K and for several particle radius. We observe again a saturation of the power exchanged between the substrate and the particle caused by the thermalization of the particle, whose temperature approaches that of the substrate as the separation is reduced. As shown in Fig.~\ref{fig3}(c), there is a maximum value of this power after the transition from the far field to near-field regime, and then the exchanged power is clearly reduced as $d$ is decreased. The corresponding particle equilibrium temperature is represented in Fig.~\ref{fig3}(d). To highlight the influence of the thermal bath, for fixed radius ($R=50\,$nm), in Figs.~\ref{fig3}(c) and \ref{fig3}(d) we take the substrate temperature as $T_1=400\,$K, while $T_3=0\,$K, $250\,$K, and $300\,$K. In the far field, we observe a strong effect of the bath on the steady-state temperature of the particle, but the exchanged power $\mathcal{P}$ is similar in the different cases (recall that the total power absorbed by the particle is always zero in the considered situations).

\section*{Discussion}

We have demonstrated the existence of a radiative saturation mechanism for near-field heat exchange in many-body systems. This saturation arises as a consequence of thermalization of the interacting bodies when the separation $d$ between them is reduced.
In contrast to the well-known saturation of heat transfer between two bodies, close to contact, resulting from the nonlocal response of the materials, the effect highlighted here exists even with purely local responses. For polar materials with planar geometry, a quadratic dependence of the temperature variation between neighboring elements is observed with respect to the separation distance. This dependence is counterbalanced by the $1/d^2$ scaling of the heat transfer coefficient and therefore, the energy flux reaches a constant value in the limit of small $d$. In metallic structures, where such a scaling does not apply, thermalization induces a saturation of the heat flux as well. 
In the considered example for Au, the saturation distance due to thermalization is one order of magnitude larger than the optical saturation distance~\cite{ChapuisPRB08} for the heat exchange between metals at fixed temperatures.

This mechanism of saturation due to thermalization can be observed in experimental measurements of radiative heat transfer in which the temperature of the active components is not completely fixed. This may be the case, for instance, of a membrane that is suspended by arms constituting a weak conductive channel for heat transport, or when a small object is attached to a cantilever whose internal temperature profile can be altered by the incoming radiative energy flux. 
The power absorbed by a particle in this more general scenario, which could represent a simplified model of a tip, can be described as $\mathcal{P}_\mathrm{abs}=\mathcal{P}_\mathrm{sus}+\mathcal{P}_\mathrm{env}+\mathcal{P}_\mathrm{ext}$, where $\mathcal{P}_\mathrm{sus}$ and $\mathcal{P}_\mathrm{env}$ account for the interaction with the substrate and the environment, respectively, and $\mathcal{P}_\mathrm{ext}$ is an external power that controls the state of the system. The term $\mathcal{P}_\mathrm{env}$ may include interactions with a bath of thermal radiation and also a conductive contribution arising from the structure supporting the particle. In the stationary state at which the absorbed power vanishes, the power $\mathcal{P}_\mathrm{ext}$ supplied to the system to maintain such a state can be used to infer the steady-state temperature of the particle and the radiative power $\mathcal{P}_\mathrm{sus}$. As we have shown here, the induced many-body thermalization can notably affect both the temperature of the particle and the power exchanged with the substrate, so it can influence experimental measurements as well. Finally, the thermalization and the associated saturation effect could be relevant for thermal management in systems with several components interacting through thermal radiation.

\section*{Methods}

\subsection*{Optical reflection and transmission coefficients, permittivities, and polarizability}

The optical reflection and transmission coefficients $\rho_j$ and $\tau_j$, respectively, of the substrate ($j=1$) and the membrane ($j=2$) are given by
\begin{equation}
\rho_1 ( k,\omega,p ) = r_1^p ( k,\omega ),\qquad
\rho_2 ( k,\omega,p ) = \frac{ r_2^p ( k,\omega ) \left( 1 - e^{2 i k_{ z  2 } \delta } \right) }{ 1 - [ r_2^p ( k,\omega ) ]^2 e^{ 2 i k_{ z  2 } \delta } },\qquad
\tau_2 ( k,\omega,p ) = \frac{ \left( 1 - [ r_2^p ( k,\omega ) ]^2 \right) e^{ i k_{ z  2 } \delta } }{ 1 - [ r_2^p ( k,\omega ) ]^2 e^{ 2 i k_{z  2 } \delta } },
\label{optical_ref_trans}
\end{equation}
and $\tau_1 ( k,\omega,p ) = 0$, since the substrate is assumed to be a semi-infinite, dissipative body. In these expressions,
$r^\text{ TE }_j = ( k_z - k_{ z  j } ) / ( k_z + k_{ z  j } )$ and
$r^\text{ TM }_j = ( \varepsilon_j k_z - k_{ z  j } ) / ( \varepsilon_j k_z + k_{ z  j } )$
are the Fresnel reflection coefficients of the vacuum-medium interfaces and 
$k_{ z  j } = \sqrt{ \omega^2 \varepsilon_j  (\omega)/c^2 - k ^2 }$
is the component of the wave vector perpendicular to the surfaces in medium $ j $ which is characterized by the dielectric permittivity $\varepsilon_ j (\omega) $.
The permittivity of SiC can be described by the Drude-Lorentz model~\cite{Palik} 
$\varepsilon(\omega)=\varepsilon_\infty (\omega^2_L-\omega^2-i\Gamma\omega)/(\omega^2_T-\omega^2-i\Gamma\omega)$, 
where $\varepsilon_\infty=6.7$ is the high frequency dielectric constant, $\omega_L=1.83\times 10^{14}\,$rad/s is the longitudinal optical frequency, $\omega_T=1.49\times 10^{14}\,$rad/s is the transverse optical frequency, and $\Gamma=8.97\times 10^{11}\,$rad/s is the damping rate. 
For Au, the permittivity here is  described by the simple Drude model
$\varepsilon(\omega)= \varepsilon_b -  \omega^2_p /( \omega^2 + i\nu\omega )$
with the background dielectric constant $ \varepsilon_b = 1 $, plasma frequency $\omega_p=1.37\times 10^{16}\,$rad/s, and electron collision frequency $\nu = 5.32\times 10^{13}\,$rad/s. 

Furthermore, in order to describe the response of the particle, we assume that its nude polarizability is given by the Clausius-Mossotti relation
$\alpha^{ (0) } (\omega) = 4 \pi R^3 [ \varepsilon(\omega) - 1 ]/[ \varepsilon(\omega) + 2 ]$, while its dressed polarizability reads
\begin{equation}
\alpha(\omega) = \alpha^{(0)}(\omega) \left( 1 - \frac{ i \omega^3 }{6 \pi c^3}  \alpha^{(0)}(\omega) \right)^{-1} .
\end{equation}

\subsection*{Correlation matrices of the substrate and bath fields}

The matrix $\mathbb{S}( \omega )$ accounting for the correlations of the substrate field can be obtained by expanding the field in plane and evanescent waves and using the correlation function of the field modes~\cite{MessinaPRA11,LatellaPRB17}. This correlation function follows from the fluctuation-dissipation theorem (\ref{FDT_incident}). Then, the matrix $\mathbb{B}( \omega ) $ describing the correlations of the thermal bath can be computed as $\mathbb{B}( \omega ) = \mathrm{ Im } \mathbb{G}( \omega ) - \mathbb{S}( \omega )$. A detailed derivation of these quantities is given in the Supplementary Information and here we give the final result:
\begin{equation}
\mathbb{S}( \omega ) = \int_0^\infty \frac{ d k }{ 8\pi }k
\begin{pmatrix}
f & 0 & 0 \\
0 & f & 0 \\
0 & 0 & g
\end{pmatrix},\qquad
\mathbb{B}( \omega ) = 
\int_0^\infty \frac{d k}{ 8\pi} k
\begin{pmatrix}
v & 0 & 0 \\
0 & v & 0 \\
0 & 0 & w 
\end{pmatrix},
\end{equation}
where
\begin{align}
f & = \Pi^\mathrm{pw} \frac{1}{2 k_z} 
\left[ \left( 1- | r_1 ^\mathrm{TE} |^2 \right) + \frac{c^2 k_z^2}{ \omega^2}\left( 1- | r_1 ^\mathrm{TM} |^2 \right) \right] 
+ \Pi^\mathrm{ew} \frac{ i }{ k_z } \left[ \mathrm{Im}\left( r_1 ^\mathrm{TE}  \right) 
- \frac{  c^2 k_z^2}{\omega^2} \mathrm{Im}\left( r_1 ^\mathrm{TM} \right) \right] e^{ i 2 k_z (d+R) } ,\\
g & = \Pi^\mathrm{pw} \frac{c^2 k^2 }{ \omega^2 k_z}\left( 1- | r_1 ^\mathrm{TM} |^2 \right)
+ \Pi^\mathrm{ew} \frac{ i 2 c^2 k^2 }{\omega^2 k_z} \mathrm{Im}\left( r_1 ^\mathrm{TM}   \right) e^{ i 2 k_z (d+R) },\\
v & =  \Pi^\mathrm{pw} \frac{1}{2 k_z} \left[ \left| 1 + r_1^\mathrm{TE} e^{ i 2 k_z (d+R)} \right|^2 
  +\frac{c^2 k_z^2}{\omega^2} \left| 1 -  r_1^\mathrm{TM} e^{ i 2 k_z (d+R)} \right|^2\right],\\
w & = \Pi^\mathrm{pw} \frac{c^2k^2}{\omega^2 k_z} \left| 1 +  r_1^\mathrm{TM} e^{ i 2 k_z (d+R) }  \right|^2 .
\end{align}

\vspace{-3mm}
\section*{Acknowledgements}

We thank A.W. Rodriguez for fruitful discussions. I.L and J.M.R. acknowledge financial support from the MICINN of the Spanish Government under Grant No. PGC2018-098373-B-I00 and from the Catalan Goverment under Grant 2017-SGR-884. J.M.R. also thanks PoreLab - Center of Excellence,  Norwegian University of Science and Technology, for financial support.
S.-A. B. acknowledges support from Heisenberg Programme of the Deutsche Forschungsgemeinschaft (DFG, German Research Foundation) under the project No. 404073166.

\section*{Author contributions statement}

I.L. and P.B.-A. initiated this study. I.L. performed the calculations with inputs from R.M., S.-A.B., J.M.R and P.B.-A.. P.B.-A. supervised the research. All authors analysed and discussed the results and reviewed the manuscript. 

\section*{Additional Information}

The authors declare no competing interests.

\clearpage

{\sffamily
\begin{center}
\textbf{\Large Supplementary Information} 
\end{center}
\vspace{3mm}
\textbf{\huge Saturation of radiative heat transfer due to \\ many-body thermalization}
\vspace{3mm}

\noindent \textbf{\large Ivan Latella$^{1,2,*}$, Riccardo Messina$^2$, Svend-Age Biehs$^3$, J. Miguel Rubi$^1$, and Philippe \\ Ben-Abdallah$^{2,*}$}

\vspace{5mm}
{
\noindent
$^1$Departament de Física de la Matèria Condensada, Universitat de Barcelona, Martí i Franquès 1, 08028 Barcelona, Spain\\
$^2$Laboratoire Charles Fabry, UMR 8501, Institut d'Optique, CNRS, Universit\'{e} Paris-Saclay, 2 Avenue Augustin Fresnel, 91127 Palaiseau Cedex, France\\
$^3$Institut f\"{u}r Physik, Carl von Ossietzky Universit\"{a}t, D-26111 Oldenburg, Germany\\
$^*$ilatella@ub.edu;pba@institutoptique.fr
}
}

\subsection*{Correlation matrices of the substrate and bath fields}

Here we obtain some expressions used in the main text. In particular, we derive the correlation matrix of the substrate field given in Eq.~(20) and an explicit expression for the matrix $\mathbb{B}(\omega)$ defining the correlations of the bath field in Eq.~(21). Below we also give explicit expressions for the factors $|\xi_i(\omega)|^2$ appearing in the exchanged powers (24) and (25). We start by considering the correlations of the substrate field.

Taking into account that the substrate field at the point $\mathbf{r}$ propagates to the right, we expand it as
\begin{equation}
\mathbf{E}^{ 1 }( \omega ) 
= \sum_{ p} \int \frac{ d^2 \mathbf{k} }{ (2\pi)^2 } 
\exp( i \mathbf{K} \cdot \mathbf{r} )
\hat{ \bm{ \epsilon } }^+( \mathbf{k} ,\omega,  p ) 
\mathcal{E}( \mathbf{k} ,\omega,  p ),
\end{equation}
where $ \mathbf{K} =(\mathbf{k},k_z) $ is the wave vector for which the component parallel to the surface is $\mathbf{k} = (k_x,k_y)$ with $k=|\mathbf{k}|$. Here,
\begin{equation}
\hat{ \bm{ \epsilon } }^\pm( \mathbf{k} ,\omega,  \mathrm{TE} ) = \frac{1}{k}(-k_y,k_x,0),\qquad 
\hat{ \bm{ \epsilon } }^\pm( \mathbf{k} ,\omega,  \mathrm{TM} ) = \frac{c}{\omega k}(\pm k_x k_z, \pm k_y k_z, -k^2)
\end{equation}
are the unit polarization vectors and $\mathcal{E}$ is the associated field mode. 
The correlation function of these field modes  is given by~\cite{MessinaPRA11-2,LatellaPRB17-2}
\begin{equation}
\begin{split}
\langle \mathcal{E}( \mathbf{k} ,\omega,  p ) \mathcal{E}^*( \mathbf{k}' ,\omega',  p' ) \rangle
&= ( 2\pi )^2 \delta( \mathbf{k} - \mathbf{k}' ) \delta( \omega - \omega' ) \delta_{pp'} 
 \frac{ \pi \hbar \omega^2 }{ \varepsilon_0 c^2} n_1(\omega)  \frac{1}{k_z}
\left[
\Pi^\mathrm{pw} \left( 1- | r_1 ^p |^2 \right)
+  \Pi^\mathrm{ew} 2 i \mathrm{Im}\left( r_1 ^p \right)
\right].
\end{split}
\end{equation}
Taking into account this correlation function and using cylindrical coordinates in which $d^2 \mathbf{k} = k dk d\phi $, the components of the correlation matrix of the substrate field can be written as
\begin{equation}
\begin{split}
\langle E_i^{ 1 }( \omega ) E_j^{ 1 *}( \omega' ) \rangle
& = \frac{ 4 \pi \hbar \omega^2 }{ \varepsilon_0 c^2} n_1(\omega) \delta( \omega - \omega' ) \\
& \times \sum_{ p } \int_0^\infty \frac{ d k }{ 8\pi }k 
\frac{1}{k_z} \left[
\Pi^\mathrm{pw} 
\left( 1- | r_1 ^p |^2 \right)
\int_0 ^{2\pi} \frac{d\phi}{2 \pi} \hat{ \epsilon }_i^+ \hat{ \epsilon }_j^{+ } 
+  \Pi^\mathrm{ew} 
2i  \mathrm{Im}\left( r_1 ^p \right) e^{ i 2 k_z (d+R)  }
\int_0 ^{2\pi} \frac{d\phi}{2 \pi} \hat{ \epsilon }_i^+ \hat{ \epsilon }_j^{-} 
\right],
\end{split}
\end{equation}
where we have made use of the properties of the polarization vectors in such a way that
$\Pi^\mathrm{pw} \hat{ \epsilon }_i^+ \hat{ \epsilon }_j^{+ *} = \Pi^\mathrm{pw} \hat{ \epsilon }_i^+ \hat{ \epsilon }_j^{+ }$
and
$\Pi^\mathrm{ew} \hat{ \epsilon }_i^+ \hat{ \epsilon }_j^{+ *} = \Pi^\mathrm{ew} \hat{ \epsilon }_i^+ \hat{ \epsilon }_j^{- }$.
Performing the angular integral in the above equation leads to the substrate filed correlation matrix~(20) of the main text, which is proportional to the matrix $\mathbb{S}$ defined by
\begin{equation}
\mathbb{S}( \omega ) = \int_0^\infty \frac{ d k }{ 8\pi }k
\begin{pmatrix}
f & 0 & 0 \\
0 & f & 0 \\
0 & 0 & g
\end{pmatrix},
\end{equation}
with
\begin{align}
f & = \Pi^\mathrm{pw} \frac{1}{2 k_z} 
\left[ \left( 1- | r_1 ^\mathrm{TE} |^2 \right) + \frac{c^2 k_z^2}{ \omega^2}\left( 1- | r_1 ^\mathrm{TM} |^2 \right) \right] 
+ \Pi^\mathrm{ew} \frac{ i }{ k_z } \left[ \mathrm{Im}\left( r_1 ^\mathrm{TE}  \right) 
- \frac{  c^2 k_z^2}{\omega^2} \mathrm{Im}\left( r_1 ^\mathrm{TM} \right) \right] e^{ i 2 k_z (d+R) } ,\\
g & = \Pi^\mathrm{pw} \frac{c^2 k^2 }{ \omega^2 k_z}\left( 1- | r_1 ^\mathrm{TM} |^2 \right)
+ \Pi^\mathrm{ew} \frac{ i 2 c^2 k^2 }{\omega^2 k_z} \mathrm{Im}\left( r_1 ^\mathrm{TM}   \right) e^{ i 2 k_z (d+R) }.
\end{align}

Furthermore, for convenience, we now express the imaginary part of the vacuum Green's function in the coincidence limit as
\begin{equation}
\mathrm{Im} \mathbb{G}^{(0)}( \omega )  
= \int_0^\infty \frac{ dk }{ 8\pi } k
\begin{pmatrix}
q & 0 & 0 \\
0 & q & 0 \\
0 & 0 & s
\end{pmatrix}, \qquad 
q= \Pi^\mathrm{pw}\frac{1}{k_z}\left( 1 +\frac{c^2 k_z^2}{\omega^2} \right),\qquad s= \Pi^\mathrm{pw} \frac{2c^2k^2}{\omega^2 k_z}. 
\end{equation}
This expression is obtained by writing the vacuum Green's function in terms of the polarization vectors in cylindrical coordinates and integrating over the angular variable. Besides, the imaginary part of the scattering Green's function in the coincident limit takes the form
\begin{equation}
\mathrm{Im} \mathbb{G}^\mathrm{(R)}( \omega ) = 
\int_0^\infty \frac{d k}{ 8\pi} k
\begin{pmatrix}
t & 0 & 0 \\
0 & t & 0 \\
0 & 0 & u 
\end{pmatrix},
\end{equation}
where 
\begin{align}
t & = \Pi^\mathrm{pw} \frac{1}{k_z} \mathrm{Re} \left[
 \left( r_1^\mathrm{TE} 
- \frac{c^2}{\omega^2} k_z^2 r_1^\mathrm{TM} \right) e^{ i 2 k_z (d+R)} 
\right]
+
\Pi^\mathrm{ew} \frac{ i }{ k_z } \left[
\mathrm{Im} ( r_1^\mathrm{TE} )  
- \frac{c^2 k_z^2}{\omega^2}  \mathrm{Im} ( r_1^\mathrm{TM} )
\right] e^{ i 2 k_z (d+R)},\\ 
u & = 
\Pi^\mathrm{pw} \frac{2 c^2 k^2}{\omega^2 k_z} 
\mathrm{Re} \left( r_1^\mathrm{TM} e^{ i 2 k_z (d+R) } \right) 
+
\Pi^\mathrm{ew} i \frac{2 c^2 k^2}{\omega^2 k_z} \mathrm{Im} ( r_1^\mathrm{TM} ) e^{ i 2 k_z (d+R) }, 
\end{align}
so that the imaginary part of the total Green's function $\mathrm{Im} \mathbb{G} ( \omega ) = \mathrm{Im} \mathbb{G}^\mathrm{(0)}( \omega ) + \mathrm{Im} \mathbb{G}^\mathrm{(R)}( \omega )$ can be readily decomposed into propagating and evanescent wave contributions as well.

Taking into account that the correlations of the bath field are given by $\mathbb{B}( \omega ) = \mathrm{ Im } \mathbb{G}( \omega ) - \mathbb{S}( \omega )$, this matrix can be written as
\begin{equation}
\mathbb{B}( \omega ) = 
\int_0^\infty \frac{d k}{ 8\pi} k
\begin{pmatrix}
v & 0 & 0 \\
0 & v & 0 \\
0 & 0 & w 
\end{pmatrix},
\end{equation}
where $v= q + t -f$ and $w = s + u - g$.
Working out these coefficients we obtain
\begin{equation}
v  =   
  \Pi^\mathrm{pw} \frac{1}{2 k_z} \left[ 
  \left| 1 + r_1^\mathrm{TE} e^{ i 2 k_z (d+R)} \right|^2 
  +\frac{c^2 k_z^2}{\omega^2} \left| 1 -  r_1^\mathrm{TM} e^{ i 2 k_z (d+R)} \right|^2\right],\qquad
w  = \Pi^\mathrm{pw} \frac{c^2k^2}{\omega^2 k_z} \left| 1 +  r_1^\mathrm{TM} e^{ i 2 k_z (d+R) }  \right|^2 ,
\end{equation}
where we observe that there is no contribution from evanescent waves.

Finally, we give an explicit expression for the factor $|\xi_i(\omega)|^2$ appearing in the spectral power, where the quantity $\xi_i(\omega)$ has been introduced in Eq.~(16) of the main text. Using the relation $|\zeta|^2=\mathrm{Re}^2(\zeta)+\mathrm{Im}^2(\zeta)$ and taking into account the expression of the scattering Green's function given in Eq.~(13) of the main text, we get
\begin{equation}
| \xi_x |^{-2} 
 = \left[ 1 + \int_0^\infty \frac{d k}{ 8\pi} k\, \frac{\omega^2}{c^2} \mathrm{Im}( \alpha a ) \right]^2 
 + \left[  \int_0^\infty \frac{d k}{ 8\pi} k\, \frac{\omega^2}{c^2}\mathrm{Re}( \alpha a ) \right]^2,
\end{equation}
with $\xi_y=\xi_x$, and $| \xi_z |^{-2}$ can be obtained from the above equation by replacing $a \to b$, the coefficients $a$ and $b$ being given by Eqs.~(13) in the main text as well.

\end{document}